# Development and Performance of the PHOT (Portable High-Speed Occultation Telescope) Systems

E.F. Young(1), L.A. Young(1), C.B.Olkin(1), K. Shoemaker(2), R.G. French(3), J. Regester(4), M.W. Buie(1)

efy@boulder.swri.edu

layoung@boulder.swri.edu

colkin@boulder.swri.edu

shoemakerlabs@gmail.com

rfrench@wellesley.edu

jregester@gmail.com

buie@boulder.swri.edu

(1) Southwest Research Institute, 1050 Walnut St, Boulder, CO 80302

(2) Shoemaker Labs, Inc., 780 Applewood Dr, Lafayette, CO 80026

(3) Dept. of Astronomy, Wellesley College, 106 Central St. Wellesley, MA 02481

(4) Greensboro Day School, 5401 Lawndale Dr, Greensboro, NC 27455

# ABSTRACT

The *PHOT* (Portable High-Speed Occultation Telescope) systems were developed for the specific purpose of observing stellar occultations by solar system objects. Stellar occultations have unique observing constraints: they may only be observable from certain parts of the globe; they often require a rapid observing cadence; and they require accurate timestamp information for each exposure. The PHOT systems consist of 14" telescopes, CCD cameras, camera mounting plates, GPS-based time standards, and data acquisition computers. The PHOT systems are similar in principle to the *POETS* systems (Portable Occultation, Eclipse and Transit Systems, described by Souza et al. 2006 and reported on by Gulbis et al. 2008), with the main differences being (a) different CCD/Cameras with slightly different specifications and (b) a stand-alone custom-built time standard used by PHOT, whereas POETS uses a commercial time-standard that is controlled from a computer. Since 2005, PHOT systems have been deployed on over two dozen occasions to sites in the US, Mexico, Chile, Namibia, South Africa, France, Austria, Switzerland, Australia and New Zealand, mounted on portable 14" telescopes or on larger stationary telescopes. Occultation light curves acquired from the 3.9-m AAT (Anglo-Australian Telescope) have produced photometric signal-to-noise ratios (SNR) of 333 per scale height for a stellar occultation by Pluto (Young et al. 2008). In this paper we describe the seven PHOT subsystems in detail (telescopes, cameras, timers and data stations) and present SNR estimates for actual and predicted occultations as functions of star brightness, telescope aperture and frame rate.

# 1. INTRODUCTION

## 1.1 Occultations

An occultation is an event in which one object passes completely behind another object. In a stellar occultation, a star is briefly obscured by a planet or other object. For solid bodies with no appreciable atmosphere or coma, such as an asteroid, an occultation results in the sudden winking out and reappearance of the stellar signal. Observations of a solid-body occultation from multiple sites results in a "raster scan" of the occulting body, yielding its size, shape, and albedo (e.g., Millis et al. 1987; Dunham et al. 1990; see our Fig. 1). For rings around giant planets or dust jets and coma around comets or centaurs, occultations measure the line-of-sight transmission along a chord cutting through the system. When a star is occulted by a planetary atmosphere, vertical variation in the atmospheric refractivity defocuses the starlight, providing temperature and density profiles of the obscuring atmosphere that could not be obtained any other way (Fig 2).

Stellar occultation campaigns have produced a steady stream of major results, including the discovery and characterization of the uranian rings (Elliot et al. 1977), the discovery of Pluto's atmosphere (Hubbard et al. 1988; Elliot et al. 1989), the super-rotating zonal winds of Titan's atmosphere (Hubbard et al. 1993; Sicardy et al. 1999), the rapidly changing column abundances of Pluto's and Triton's atmospheres (Olkin et al. 1997; Elliot et al. 2000, oung et al. 2008), the characterization of gravity waves in the atmospheres of Titan and the gas giants (French and Geirasch 1974; Sicardy et al. 1999), and the refinement of sizes and positions for tens of asteroids per year.

[Editor: Place FIG. 1 here]

[Editor: Place FIG. 2 here]

Because solar system objects have well-known orbital motions, improved time resolution of a stellar occultation observation translates into better spatial resolution of the occulting object. For objects beyond the asteroid belt, the star's relative velocity in the object's sky plane is typically around 20 km sec$^{-1}$, dominated by the Earth's orbital motion, but can be much less if the occultation occurs near the stationary point of the object's orbit. The two main properties of the PHOT design (portability and good SNR at high frame rates) are driven by the need to observe from sites around the world and at a cadence that oversamples the duration of typical occultation events.

The remainder of this paper is laid out as follows: §1.2 describes the design requirements for a portable occultation system, §2 describes the actual components of the *PHOT* systems (including cameras, timing systems, data acquisition systems, the telescopes themselves and provisions for packing and transporting the *PHOT* systems), and §3 compares the predicted performance of the *PHOT* systems to actual performance observed during past occultations.

## 1.2  PHOT Design Requirements

### *1.2.1 Frame Rates*

What frame rates are needed to characterize the size of an object or an atmosphere's vertical profile? It is reasonable to assume an upper limit for the relative velocity between the occulter and the occulted star of 20 km sec$^{-1}$; only a subset of asteroids or near-earth objects will be faster. Given that relative velocity, a frame rate of 20 Hz will take data at 1 km intervals.

While it would be useful to characterize many phenomena at the 1 km level, other factors can impact the resolution at this level. The angular diameter of the star can be a limiting factor: for the 28 Sgr occultation by Titan, the effective stellar size was estimated to be 20 km. For fainter occultation stars, the stellar size is more typically around 1 km.

The Fresnel diffraction limit is a second limiting factor.

$$Fresnel\ Limit = \sqrt{\lambda D/2} \qquad (1)$$

where $\lambda$ is the wavelength of the observation and $D$ is the distance to the occulter. For visible wavelength events ($\lambda$ = 700 nm) at Jupiter's average distance from the Earth, the Fresnel scale is 0.7 km. For infrared (2.2 $\mu$m) events at 30 AU, the Fresnel scale is 4.45 km.

The scale heights of the atmospheres of Jupiter, Uranus, Titan and Pluto range from 20 to 60 km. A frame rate of 2 Hz will oversample the scale heights, and faster frame rates will reveal density and temperature perturbations on the scales of a few km. Accordingly, we have chosen frame rates up to 20 Hz as a *PHOT* design goal, which means that the *PHOT* cameras must have dead time between frames that is small compared to 0.05 s. Near-zero dead time between exposures is routinely achieved with frame transfer (FT) CCDs. With FT CCDs, the dead time between exposures is the time required to transfer charge from the active region of the CCD to the storage region, typically a few microseconds per parallel shift or a few milliseconds per exposure. The storage region can be read out without disturbing the active region. Provided the read-out time is less than the requested exposure time, a sequence of frames can be obtained with only a few milliseconds of dead time between frames.

The read-out time is often inversely correlated with the read noise: at the time the PHOT systems were assembled, CCDs with 5 - 10 MHz read-out rates typically had read noises in the 10 e-/read range, while slower rates (e.g., 100 KHz) achieved read noises in neighborhoods of 1 to 3 e- per read. There are ways to improve the overall frame rate and still reap the benefits of low read noise system. Consider 100 KHz as a lower limit: at 100 KHz, a 512 x 512 frame read-out would take about 2.5 s, much longer than the target exposure time of 0.1 to 0.05 s. In practice, however, one can usually reduce the number of read-out operations per frame by reading subframes around the occultation star and binning pixels in hardware. The disadvantage of reading small subframes is the reduced number of on-chip stars that might be used as photometric standards. Nevertheless, a 128 x 128 subframe, binned 2 x 2, takes just over 0.04 s per frame to read at 100 KHz and is compatible with a 20 Hz frame transfer system.

Hardware binning is appropriate when the CCD's plate scale vastly oversamples the seeing. The plate scale is inversely proportional to the telescope's focal length. The shortest focal lengths we have used in four years of PHOT deployments has been 3750 mm on our 14-inch Schmitt-Cassegrain telescopes. This focal length produces a plate scale of 0.055 arcsec/$\mu$m (or 0.7 arcsec/pixel for our 13 $\mu$m pixels). This plate scale oversamples the seeing resolution from most (but not all) sites; we would consider binning (2 x 2) only if the seeing were worse that 2 arcsec or if there were a need for a wide field of view (FOV) at fast frame rates. If the seeing conditions vary substantially from frame to frame, one might extract a photometrically better lightcurve using PSF (Point Spread Function) fitting instead of aperture photometry. In these cases, there is a benefit to oversampling the PSF as part of the PSF fitting process.

Software binning (after individual pixels have been read out, as opposed to on-chip binning in hardware) is never appropriate. Like hardware binning, it reduces the the spatial resolution, but unlike binning in hardware, it provides no improvements in read-out time or higher signal-to-noise ratios (SNR).

## 1.2.2 Noise Sources

There are many more faint stars than bright ones, and many more occultations of faint stars than bright ones. The practical faint-event limit depends on the SNR that can be obtained during an event, which in turn depends on brightness of the occulted star, the occulting object, the maximum acceptable exposure time, the read noise and the noise from combined background sources.

$$SNR = \frac{S}{\sqrt{rn^2 + S + O + BG}} \qquad (2)$$

where the numerator, S, is the signal (counts from the occulted star in e-) and the denominator is the noise, which is the quadratic sum of variances due to read noise ($rn^2$) and photon shot noise from the source (S), the occulter (O) and background sources (BG). Equation (2) assumes that the noise sources are uncorrelated and that all noise terms are in units of electrons. Dark current has never been a significant noise source in any of our occultation campaigns, due to the short exposure times that have been used. In addition to terms in the denominator of (2), other systematic noise sources can corrupt a lightcurve, like variable clouds or scintillation. We always ratio the flux from the occulted star to one or more on-chip comparison stars. The widest FOV used to date with the PHOT systems has been 6 arcminutes (with the 3750 mm focal-length portable 14-inch telescopes), small enough to insure that the entire FOV is affected similarly by cloud cover, scintillation and variable seeing.

Although the *variance* in counts from the occulted object *O* (which is equal to the *counts* from *O*) might be considered to be part of the background, we have explicitly called it out as a separate noise source to emphasize that the photon shot noise from the occulting object often dominates all other noise sources.

*Read Noise vs. Photon Shot Noise*

When the occulting object and the occulted star are both faint, the dominant noise source is likely to be read noise. This is especially true if the exposure times are short. Since frame rates translate to spatial resolution, exposure times are nearly always as short as possible, up to resolutions corresponding to the Fresnel limit. These objectives dictate the main characteristics of a good occultation detector: high quantum efficiency (QE), high frame rates, low read noise, and zero dead time between exposures.

For faint, read noise-limited events, low read noise and high QE are the most essential CCD characteristics. For example, consider two cameras that only differ in their read noise values, one at 3 e- per read per pixel, the other at 10 e-. The improvement in SNR due to lower read noise is equivalent to a factor of ten in exposure times or in occultation star brightness. The camera with 3 e- read-noise will be able to sense approximately four times as many events as the 10 e- camera at a given SNR limit.

*The Case for EMCCDs*

CCD vendors have achieved essentially zero read noise cameras with *electron multiplication CCDs* (EMCCDs). These devices have extended serial registers that amplify the signal from a pixel before it is read by an ADC. Even if the ADC has a read noise of 10 or 100 e-, the effective read noise is nearly zero after dividing by the serial register gain factor. The downside of these devices is that other noise sources are amplified by the serial register by 40% (Robbins & Hadwen 2003), so using an EMCCD is roughly equivalent to halving the size of the aperture and eliminating read noise from the detector. For very faint events, the EMCCD mode can be advantageous, but for cases in which the occulting object is as bright as Pluto, the read-noise is overwhelmed by photon shot noise of the sources, even for 14-inch apertures. In §3.1 we present a signal-to-noise calculator and compare the predicted SNR for EMCCD vs. conventional CCD modes. EMCCDs also have lower dynamic ranges than conventional read-out CCDs. This may not be a factor in occultation observations at high frame rates (e.g., 20 Hz) where all the sources in the frame are read-noise limited, but could potentially prevent the use of a bright comparison star in some cases.

### 1.2.3 Detector Sensitivity

Detector sensitivity, or quantum efficiency, is a simple design goal to articulate: we want the QE to be as close to 100% as possible. High QEs are typically achieved by using thinned, back illuminated CCDs. At the time that the original PHOT cameras were purchased from Roper Scientific (late 2004), the extra step of thinning and back-illumination raised the company's CCD's QEs from around 40% to 90% over the 0.45 $\mu$m - 0.75 $\mu$m wavelength range.

[Editor: Place FIG. 3 here]

### 1.2.4 Dark Current

Dark current is not a major noise source for most occultations because the exposure times are generally short. Nevertheless, cooled CCDs are available and should be part of the purchase decision, since dark current generally doubles for every 7°C rise in detector temperature. The nominal dark current for the PhotonMax cameras at -55°C is 0.008 electrons per pixel per second.

Both the MicroMax and PhotonMax CCDs are thermoelectrically cooled, to -45°C and -80°C, respectively. Liquid $N_2$ systems are also available but probably not necessary for occultation observations. One concern with a thermoelectrically cooled system is power consumption, which can be several hundred watts during the initial few minutes of cool-down. The *PHOT* systems have been run from a car batteries, but a powerful inverter (capable of more than 800 W at 110 V) was needed to run the CCD, telescope, laptop and timer.

*1.2.5 Accurate Times and Triggered Exposures*

One of the most critical aspects of occultation observations is the need for accurate knowledge of the absolute beginning and end times of each exposure. Occultations are almost always observed from multiple sites, since multiple chords break the degeneracy between uncertainties in the geometry of the event and sought-after physical parameters. Separate data sets need accurate absolute timing if they are to be combined in a useful way. Our requirement for absolute timing accuracy is 500 $\mu$s, corresponding to about 1% of the duration of a 0.05 s exposure.

The need for accurate timestamps leads to the requirement that the camera can be triggered to take each exposure, from an external TTL-level signal, for example. Many cameras can be triggered from a host computer, but it can difficult to accurately set, verify and maintain a laptop's internal clock in the field at the millisecond level. For this reason we require a camera for which each individual exposure can be triggered from a signal with an accurate time pedigree.

We generate triggers with *AstroTimers*, stand-alone, custom-designed and built time standards. These GPS-based units are described in more detail in §2.2.

### 1.2.6 Summary of Requirements

[Editor: Place Table 1 here]

## 2. TECHNICAL DESCRIPTION OF THE PHOT SYSTEMS

### 2.1 Cameras

The PHOT systems use two different model cameras from Roper Instruments. Four of the PHOT cameras use *MicroMAX 512BFT* units. Three of the PHOT cameras use *PhotonMAX 512B* units. Although the PhotonMAX cameras were purchased because the MicroMAX units were no longer available, the two models have distinctive strengths which make them appropriate for different types of events.

*Common Characteristics*

Both cameras have

- Back-illuminated CCDs and high QEs (Fig. 3)
- TE (thermoelectric) cooling
- 512 x 512 pixels

- Frame transfer operation
- Capability of using external triggers
- Subframe readout modes
- Hardware binning

The MicroMAX cameras have two ADCs that can be selected via the data acquisition application. These run at 100 KHz and 5 MHz, with read noises of 3.1 and 11 e-, respectively (the nominal read noise at 100 KHz is 4 e-, but we measured lower read noises in our lab tests). The MicroMAX camera is connected to a data acquisition laptop via a USB 2.0 interface. This common interface allows us to use nearly any laptop on the market that can run Windows XP and WinView, the data acquisition application from Roper Scientific. The MicroMax cameras are about the size and weight of two bricks (4.5" x 4.5" x 8"; 7.5 lbs). They are run from a separate electronics box that is about the size of a toaster oven and weighs about 20 lbs. The MicroMAX cameras are thermoelectrically cooled to about -45°C. There is a vacuum around the CCD assembly that occasionally needs to be pumped down.

The PhotonMAX cameras have two conventional and two EMCCD ADC modes (four total). The conventional modes are at 1 and 5 MHz, with corresponding read noises of 7 and 11 e-, respectively. The electron multiplication modes are at 5 and 10 MHz, with negligible effective read noises. The PhotonMAX cameras have a factory-sealed vacuum that cannot be pumped (and is supported by a lifetime warranty). They are thermoelectrically cooled to -80°C. The PhotonMAX cameras are slightly bigger than the MicroMAX CCD head units but have no separate electronics boxes. The PhotonMAX data rates are too fast for USB 2.0; they interface to a data acquisition computer via a half-height PCI-Card. The PHOT systems use Dell D620 laptops with docks to host the half-height cards.

[Editor: Place FIG. 4 here]

## 2.2 Timers

We designed and built two generations of *AstroTimers*. Both are based on GPS chips that were specifically intended to provide accurate times. The GPS chips provide the following:

- NMEA strings (*National Marine Electronics Assoc.*) with latitude, longitude, altitude, time, number satellites being tracked and dilution of precision (a measure of the accuracy of the position/time solution)

- A 1-PPS (pulse per second) output, consisting of a TTL (transistor-to-transistor level) signal with a rising edge every second. The nominal accuracy of this PPS (pulse-per-second) signal is 30 - 70 ns, according to NavSync (the GPS chip manufacturer) literature.

- A variable frequency output, synched to the PPS signal, user-selectable from 10 KHz to 10 MHz. We use the GPS chipset's 10 KHz output to generate user-definable signal trains. The AstroTimers' pulse counting subroutines impose a delay of 1.3 ms to these signal trains, but this delay is a measured quantity that varies by less than 0.1 ms.

Both AstroTimers use two or more single board computers (SBCs) from Vesta Technology, Inc. The main purpose of the SBC is to generate triggers in the form of a signal train from the 10

KHz output of the GPS chip. In the first generation AstroTimers, the user specifies a start time, a pulse interval and pulse duration, and number of pulses to generate. One SBC is dedicated to counting 10 KHz pulses and generating the requested sequence of pulses. Two other SBCs are dedicated to (a) providing a user interface via a 16-character keyboard and (b) updating a display of the UT time.

The second generation AstroTimers are simpler than the first generation units. They lack the ability to pre-program the start time. Instead, the pulse train begins on an integer second shortly after the user flips a switch. The exact start time is displayed when the sequence of pulses begins. The second generation AstroTimers use a single SBC to count 10 KHz pulses and generate the signal train as well as providing a user interface via push buttons and toggle switches.

The second generation AstroTimers use a CW12-TIM GPS receiver model manufactured by NavSync. This module is a small (40 x 60 x 10 mm) board. The rms error on the 1 PPS output is rated at 30 ns. The nominal time for a typical cold start is 45 seconds.

[Editor: Place FIG. 5 here]

The ability to generate a specific number of triggers (and no more) has turned out to be an important feature. The data collection software can also be programmed to collect a certain number of images. If, after an AstroTimer has generated $N$ pulses, the data collection software reports that fewer than $N$ exposures were triggered, then we know that some triggers were missed. In practice we have never lost triggers in any deployments.

Both versions of AstroTimers run off of 12 VDC. Both have SMA connectors to GPS antennas. Both have BNC connectors to (a) PPS output and (b) the user-specified pulse train. The PPS output comes directly from the GPS chip. It is intended to provide some basic triggering capability in the event of a malfunction of an SBC.

## 2.3 Telescopes and Mounts

### 2.3.1 The Portable 14-Inch Telescopes

The choice of 14-inch telescopes was a compromise between the opposing goals of (a) maximizing the signal-to-noise ratio during an occultation and (b) maintaining the portability of the telescope. The PHOT systems include four 14-inch Meade Schmidt-Cassegrain telescopes. These telescopes have a focal length of 3750 mm in a relatively compact form. The combined weight of the tripod, fork mount and optical tube assembly is about 190 lb.

At the time that these telescopes were purchased, there were telescopes of similar aperture being sold by competing vendors. Except for the Alt-Az Meade telescopes, all comparable telescopes used German equatorial mounts and therefore unacceptable for occultation work. German equatorial mounts cannot track an object across its hour angle = 0 position without "flipping the telescope over the pier" first. The ability the telescope to track objects across the sky without interruption was an essential requirement for occultation observations. The chief disadvantage of Alt-Az systems, the need for an image rotator for long exposures, is not relevant to occultation work. (Even with image rotation, it is a simple matter to identify the approximate positions of the occultation star and the comparison stars with – sometimes we use known offsets (and known rota-tion angles) from the brightest star in the frame. Once the approximate positions are known, our photometry pipeline determines sub-pixel centroids for stars and extracts their fluxes.)

The Alt-Az drive of the Meade 14-inch telescopes was the deciding factor in choosing a portable telescope. In addition, the Meade telescopes have GO-TO electronics that vastly simplifies finding the occultation field. Finally, the telescopes can run on 8 C-cell batteries or on 12 VDC.

### *2.3.2 PHOT Systems on Large Telescopes*

Most of the PHOT deployments have been used on fixed telescopes at existing observatories. Although many facilities already have visible wavelength cameras, the advantages of the PHOT cameras (low read noise, high QE, virtually no dead time, high frame rates, precise time stamping) have proved to be worth the effort in transporting and mounting the cameras on a fixed telescopes.

We have built adapters for both types of cameras that add a short 2-inch outer diameter tube to the front of the CCD, centered on the active region of the CCD. This "snout" adapter lets us easily mount the cameras to telescopes that already accept 2-inch eyepieces. We have also machined adapter plates to interface to many different telescopes, such as the AAT (Fig. 6).

[Editor: Place FIG. 6 here]

## 2.4    Data Acquisition

Roper Scientific distributes a application, *WinView*, to run their MicroMax and PhotonMax cameras. They also distribute a cross platform library, *PVCam*, for development of data acquisition applications on Macintosh, Windows and Linux platforms. While it would be nice to write a program tailored to occultation observations, we have found WinView to be adequate. It allows the user to specify hardware binning, sub-frame location and hardware triggering, all of which are necessary for our occultation runs.

We use laptops running Windows XP Professional as data acquisition stations. The MicroMax cameras can be operated from virtually any PC with a USB 2.0 port. Because the PhotonMax data rate has a significantly higher ceiling, those cameras interface to the host computer via a half-height PCI card. We run the PhotonMax cameras from Dell Latitude D620 laptops connected to docking stations which have the Roper PCI cards installed.

The data acquisition laptops are typically within 15 feet of the camera or electronics box, constrained by the length of a USB or data cable. On large telescopes, we often have to fix the acquisition laptop to the telescope structure and provide access to the laptop from a distant control room. There are several ways to control a laptop remotely; we most often use VNC for this purpose. The acquisition laptop is either connected to a network via ethernet or WiFi, or a long ethernet cable is used to directly connect the control room computer to the laptop. The time lag of the VNC connection has not been an obstacle in any of our observing campaigns.

## 2.5  Packing and Transport

Telescopes are transported in wheeled, foam-lined fiberglass cases from JMI (Jim's Mobile Incorporated, Golden, CO) that are sold specifically for the 14-inch Meade telescopes. These cases are sufficient for transport within a car, but not for shipping. In 2006 we had two wooden cases built by Crating Technologies (Longmont, CO) to contain the JMI cases and tripods. (Wood used in internationally shipped crates has to be certified for that purpose.)

The total round-trip cost of shipping a telescope (including carnets or other customs documents, shipping containers, etc.) to distant international destinations has been close to $3000, or about half the cost of the telescope itself. For that reason we have decided to leave one telescope in Tasmania for future southern hemisphere events.

[Editor: Place Table 2 here]

[Editor: Place Table 3 here]

The PHOT cameras, laptops and AstroTimers can be checked as luggage in padded cases made for us by Atlas case (Denver, CO). We use Pelican-Hardigg cases (model iM2750), with foam cut-outs for cameras, laptops, docks (for the PhotonMax cameras), AstroTimers and cables. Weight of the case alone is about 18 lbs; the loaded cases are close to 48 lbs, just under the 50 lb oversize limit currently imposed by many airlines.

[Editor: Place FIG. 7 here]

## 3. EVALUATION OF THE PHOT SYSTEMS

In this section we try to address two questions: what is the predicted performance of the PHOT systems, and how have the PHOT systems lived up to these predictions?

### 3.1 Predicted SNR

We have written signal-to-noise estimators to help compare cameras with different characteristics, to compare binning factors and to decide how fast a cadence will be possible for an event of a given brightness. Fig. 8 shows predicted SNRs given the following assumed parameters:

- Sky $V_{mag}$ = 22.5

- Point source Full-Width at Half-Maximum (FWHM) = 2.0 arcsec

- Telescope aperture: 14 inches

- Clear aperture: 87.6% (a 4.9 inch secondary)

- Telescope transmission at 750 nm: 83% (for Meade's UHTC optics)

- Read noise for the MicroMAX 512BFT camera: 3.1 e-

- Read noise for the PhotonMAX 512B EMCCD camera: effectively zero

- PhotonMAX 512B "excess noise factor" in EMCCD-mode for noise sources besides read noise: 1.4.

[Editor: Place FIG. 8 here]

## 3.2  Performance from Past Deployments

This section compares predicted vs. observed performance for two previous Pluto occultations, one which used a MicroMax 512BFT 14-in telescope (from a 31-JUL-2007 event observed from Tasmania) and another which used a MicroMax 512BFT camera mounted on the 4-m AAT telescope (from the 12-JUN-2006 event observed from Siding Springs). These two cases are end members (in aperture) among the roughly two-dozen deployments that have been undertaken to date with *PHOT* cameras.

The 31-JUL-2007 event was observed from Musselroe Bay in Tasmania. This event illustrates that the value of a portable telescope can outweigh the effort and expense associated with its shipment and operation. This telescope provided an important chord across Pluto's shadow path in a region that did not have fixed telescopes of a comparable aperture, but more significantly, the observers (in consultation with the local branch of the Australian Bureau of Meteorology) were able to pick a location that had the best prediction for clear skies. In contrast, observers at the Mt Canopus Observatory in Hobart had significant cloud cover through most of the event. The occulted star was labelled P495.3 on our internal watch lists; it was a moderately fast event (sky plane velocity of 16.4 km/sec) and relatively bright (V-mag of 13.25). The predicted vs. observed performance of the PHOT system on a 14-inch telescope is shown in Fig. 9.

The 12-JUN-2006 event was observed from the 3.9-m AAT telescope in Siding Spring, Australia. The unfiltered magnitude of the occultation star (P384.2 in McDonald and Elliot 2000a,b) was measured to be 0.83 magnitudes fainter than Pluto on the night of the event (about a V-mag of 15), with a sky plane velocity of 24.0 km/sec. The predicted vs. observed performance of the PHOT system on a 3.9-m telescope is shown in Fig. 11.

[Editor: Place FIG. 9 here]

[Editor: Place FIG. 10 here]

[Editor: Place FIG. 11 here]

## 4. SUMMARY AND CONCLUSIONS

The PHOT systems requirements (§1.2.6) were originally intended to enable observations of stellar occultations by Pluto or Triton from sites around the world. A PHOT system was first deployed to South America in 2005 observe a stellar occultation by Charon, with excellent results. Since then PHOT systems have been deployed over two dozen times, on telescopes ranging from 14-inch Meades to the 200-inch telescope at Palomar. In 2006, a PHOT system mounted on the AAT obtained the best (highest SNR) Pluto occultation light curve to date, providing evidence of gravity waves in Pluto's atmosphere, the relative lengths of dynamical vs. radiative timescales, the bulk extent of Pluto's atmosphere, evidence for a non-isothermal profile in Pluto's upper atmosphere, and evidence for a temperature inversion (as opposed to a haze layer) to explain the drop in transmission through Pluto's lower atmosphere (Young et al. 2008).

The performance of the PHOT systems has been well characterized (Fig. 9 and Fig. 11). We are able to predict the SNR for an event with good accuracy and choose optimum frame rates and binning factors in advance.

In general, the PHOT systems meet or exceed the requirements outlined in §1.2.6. They have been deployed to international destinations (e.g., to Tasmania or Namibia) with as little as three weeks lead time (if a 14-inch Meade telescope has to be shipped) or two days lead time if only the cameras, AstroTimers, and laptops are necessary. We are currently building a prototype of an airline-checkable 22 inch/100 lb telescope. This telescope would reduce the lead time for deployment to most of the globe to 2 days, albeit with possible charges for excess baggage.

## *Acknowledgements*

The equipment and work described in this paper were supported by the NSF's Major Research Instrumentation Program and NASA's Planetary Astronomy Program. The authors wish to acknowledge an anonymous reviewer for valuable comments and corrections.

**Table 1.**

| Feature | Goal | Performance of PHOT |
|---|---|---|
| Frame Rate | 1 – 20 Hz | 10 Hz (routine), 20 Hz w. a 64x64 subframe |
| Dead Time | < 10% of frame rate | < 4% of max frame rate w. frame transfer CCD |
| Field of View | 1 – 20 arcmin | 6 arcmin w. the 14-in Meade telescopes, 0.4 arcmin from the 4-m AAT |
| Plate Scale | 0.5 – 1.2 arcsec/pixel | 0.75 arcsec/pixel w. 14-in Meades |
| Timing Knowledge | < 1 ms | < 1 µs for 1 PPS, < 1 ms for user-defined pulse trains |
| Deployability | Worldwide, < 3 weeks lead time | 2 weeks typical |
| Ease of Setup | Observe < 1 hr after first star acquisition | 30 min setup typical |
| Sensitivity Pluto Atm. | SNR = 16 at 5 Hz | For $V_{OCC}$ = 15.5, measured SNR per exposure = >20 at 5 Hz with 14-in Meades or >50 at 10 Hz with 4-m AAT |

**Table 2. Size and Weight of Components**

| Component | Size (packed, inches) | Weight (lbs) |
|---|---|---|
| Detector | 5 x 5 x 8 | 8 |
| CCD Controller | 6 x 9 x 13 | 20 |
| Laptop | 1 x 9 x 11 | 7 |
| Dock | 3 x 19 x 16 | 8 |
| GPS Time Standard | 2 x 6 x 9 | 3 |
| OTA + Mount | 17 x 26 x 38 | 160 |
| Tripod | 48 x 12 x 12 | 50 |
| Field Kit (Tools & Parts) | 6 x 8 x 14 | 25 |
| Telescope Case (JMI) | 20 x 25 x 48 | 45 |
| Camera & Laptop Case | 18 x 20 x 24 | 12 |
|  |  |  |
|  |  |  |
| **TOTAL** |  | **338** |

**Table 3.**     **Component Power Requirements**

| Component | Power Requirements | Power Source(s) |
|---|---|---|
| Camera | 300 W[1] | 120 - 240 VAC or Car[2] |
| Laptop (with Dock) | 90 W | 120 - 240 VAC or Car[2] |
| Astrotimer | 10 W | 12 VDC |
| Telescope | 1 W | 12 VDC or 8 C-cell batteries |

[1] The power drawn by the CCD during the initial cool-down phase can exceed 300 W. When running the MicroMax CCDs from car-battery, an 800 W (or greater) inverter was required during the cool-down period.

[2] "Car" means "120 VAC inverter powered from a car battery/cigarette lighter." On past observing runs, the telescope, camera, laptop and astrotimer were all powered from a car (with the engine running) and an 800 W inverter.

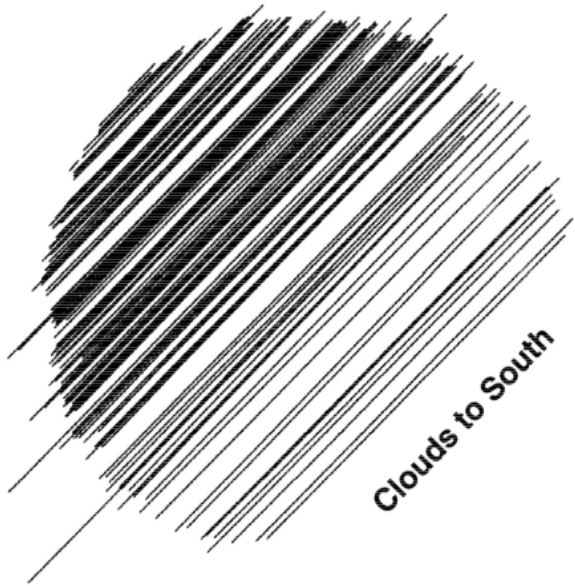

**FIG. 1.**

Diagram of a stellar occultation by a airless body. Different observing sites probe different parts of the occulting body's shadow as it passes over the Earth. Combining multiple observations produces a raster scan of the object, yielding information on its size and shape. A single chord can only provide a lower limit to the object's radius. This example shows the results of an occultation by Pallas (Dunham et al. 1990).

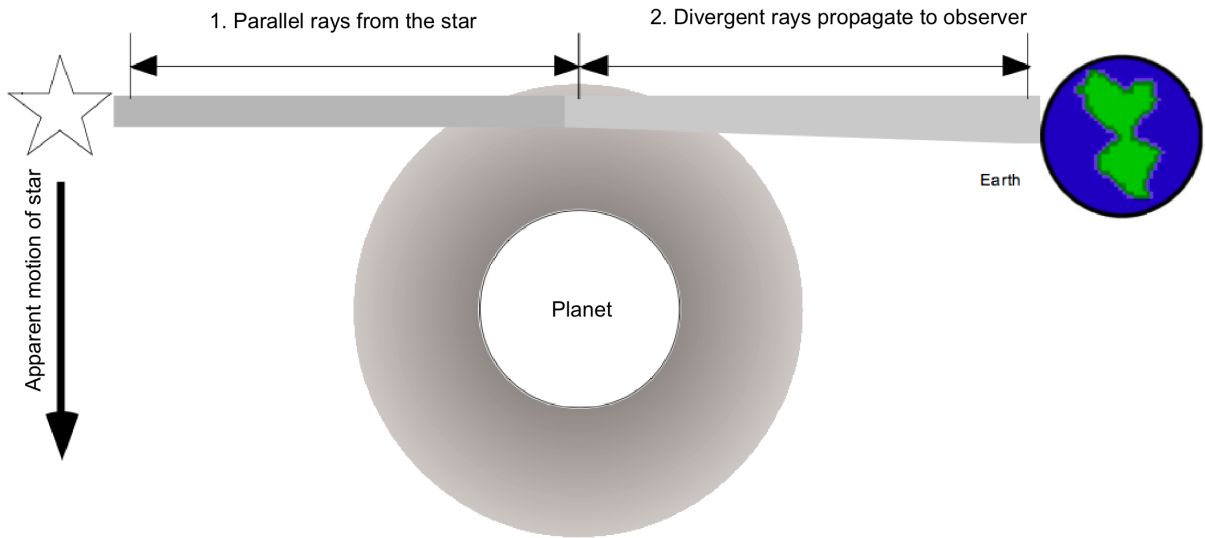

**FIG. 2.**

Diagram of a stellar occultation by a planetary atmosphere. The star appears to move behind an object due to the relative motions of the Earth, a star, and the occulting object. The star fades during ingress and reappears during egress. Even a clear atmosphere will attenuate starlight before the surface of the occulter is encountered. Because atmospheres are generally thicker at lower altitudes, rays from the star are refracted more if they impact the occulter's atmosphere near the surface. The attenuation due to differential refraction is often more significant than attenuation due to atmospheric opacity by hazes and aerosols. Because of differential refraction, occultations in visible and near-IR wavelengths typically probe down to µbar pressure levels.

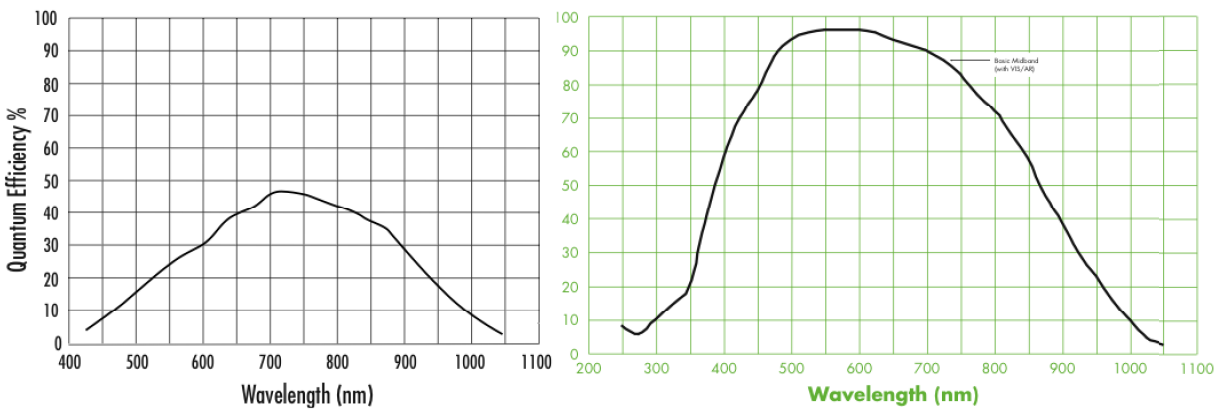

**FIG. 3.** Quantum Efficiency (QE) curves for the Roper MicroMax 512FT (*left panel*) and MicroMax 512BFT (*right panel*). The BFT designation means that the detector is a thinned, back-illuminated, frame-transfer CCD. The advantage over a non-thinned, non-back-illuminated CCD is more than a factor of 2x increase in QE.

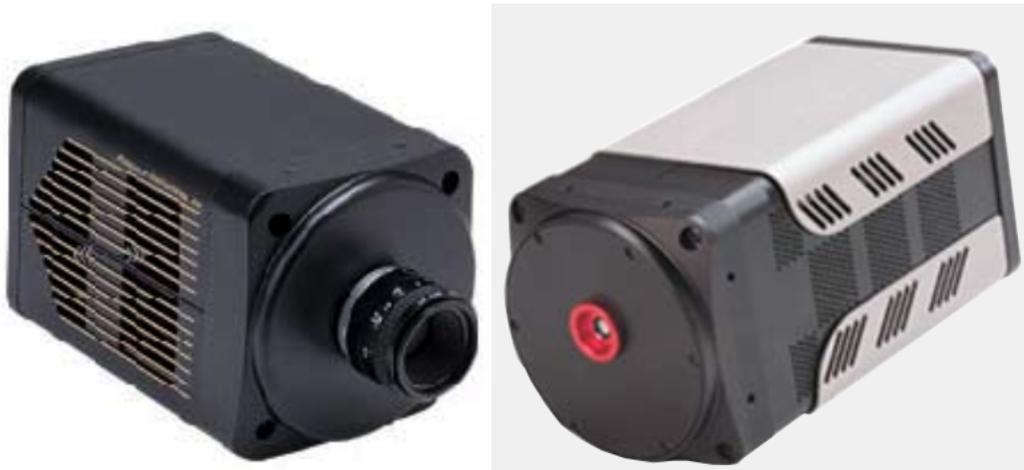

**FIG. 4.** The MicroMax 512BFT camera (*left*) and the PhotonMax 512B camera.

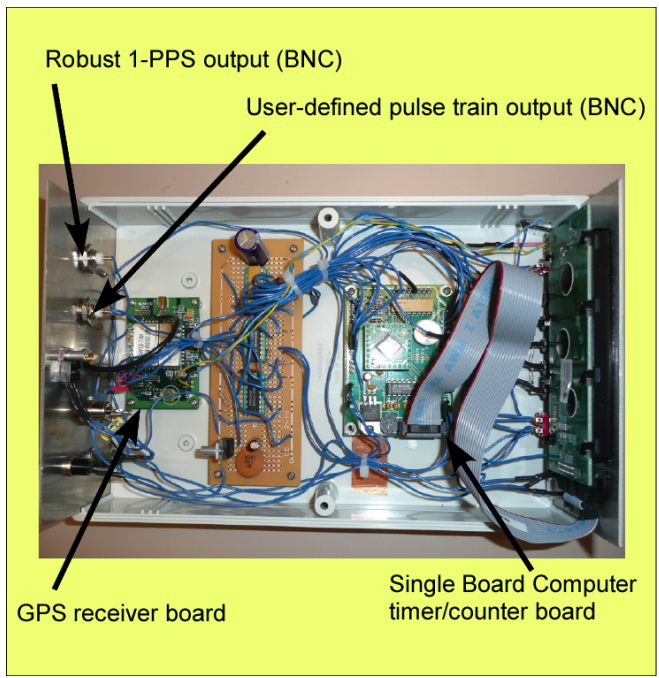

**FIG. 5.** Inside the GPS-based AstroTimer. Outside dimensions are 6" x 10" x 3"

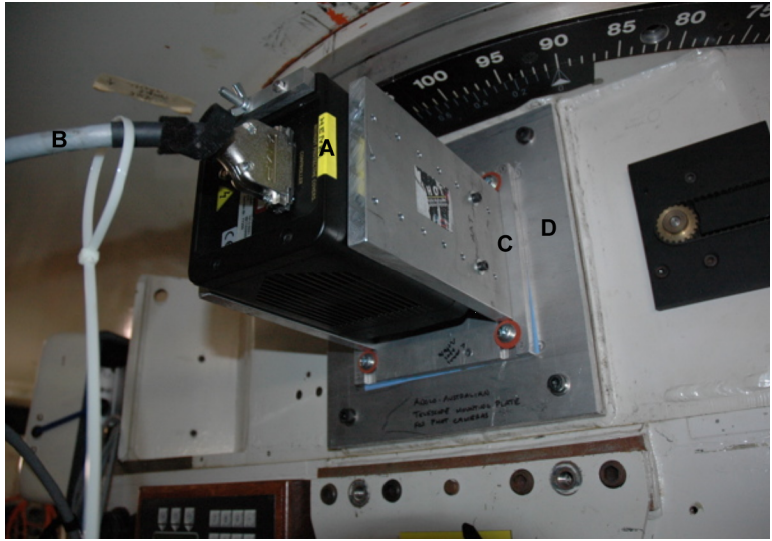

**FIG. 6.**

A PHOT camera (MicroMax 512BFT) mounted on the auxiliary port of the 3.9-m Anglo-Australian Telescope (AAT). Shown here are the MicroMax CCD head unit (A), the cable (B) connecting it to the electronics box, the aluminum stand-offs to which the camera is bolted (C), and an adapter plate (D) with a bolt hole pattern to mate the stand-off unit to the telescope.

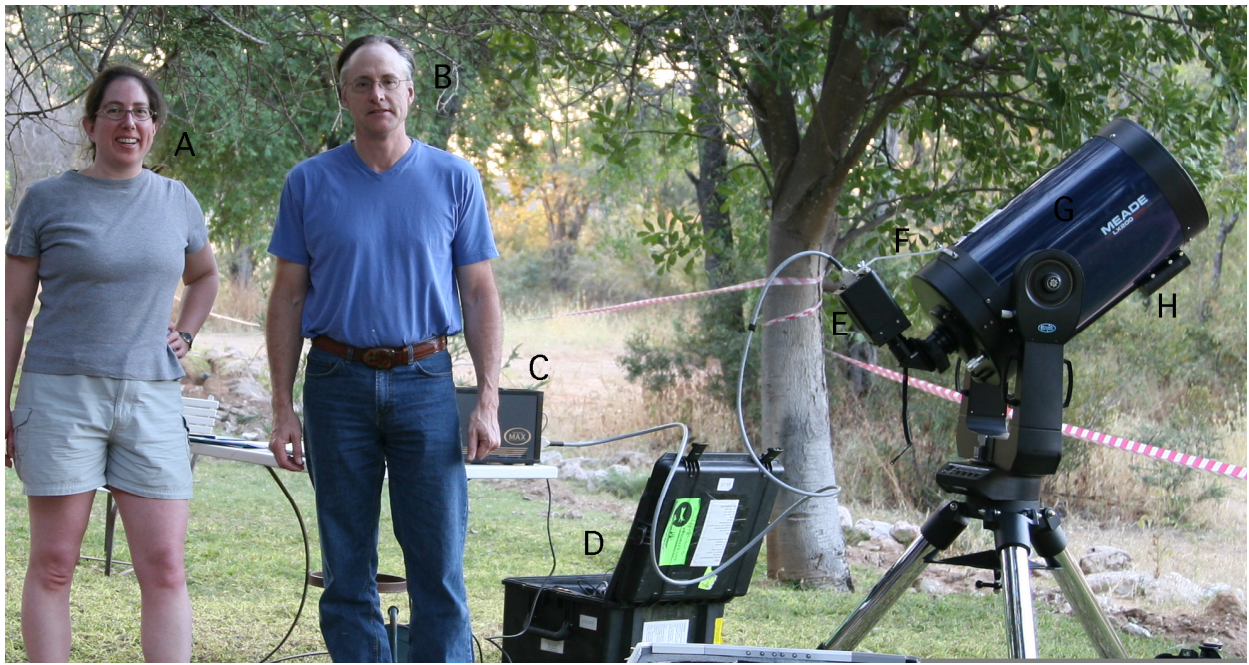

**FIG. 7.**

A *PHOT* system deployed in Namibia in April 2009. Components of the PHOT system include astronomers Leslie Young and Marc Buie (A,B), the MicroMax camera electronics box (C), the Hardigg foam case for the MicroMax camera and the AstroTimer (D), the MicroMax CCD head unit (E), a strain relief bracket for the CCD (F), a 14" Meade telescope (G), and third-party counter-weights (H).

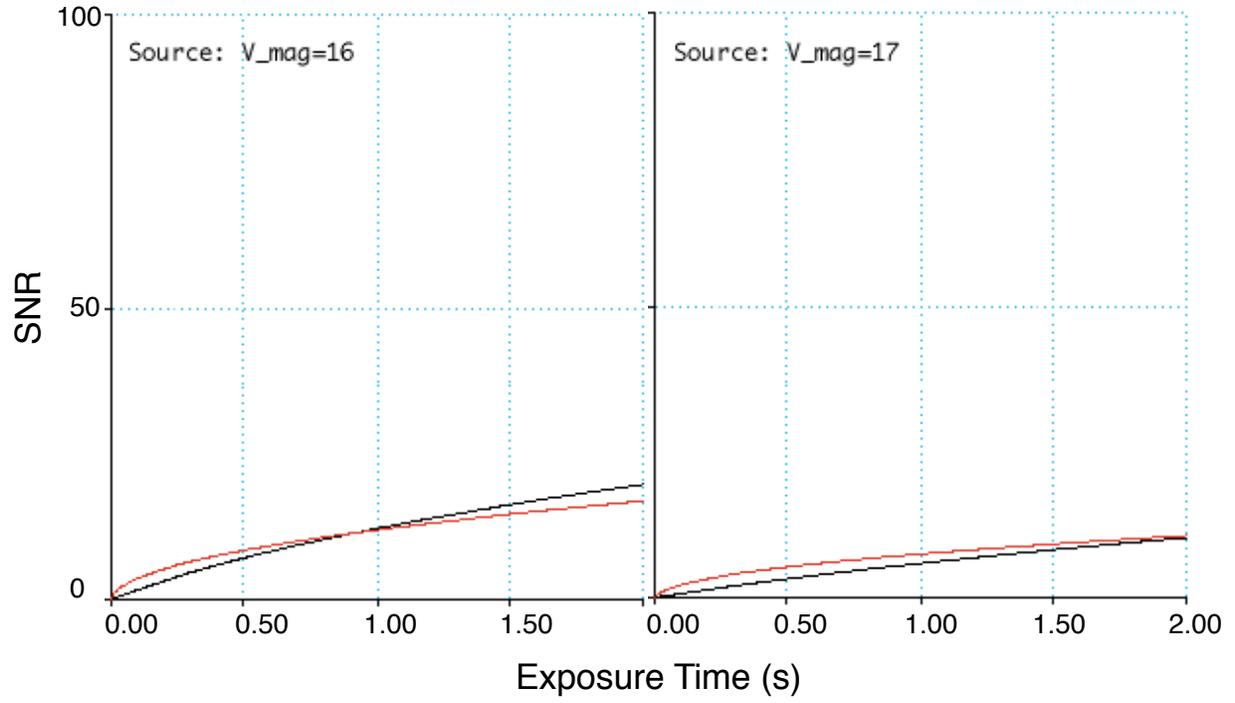
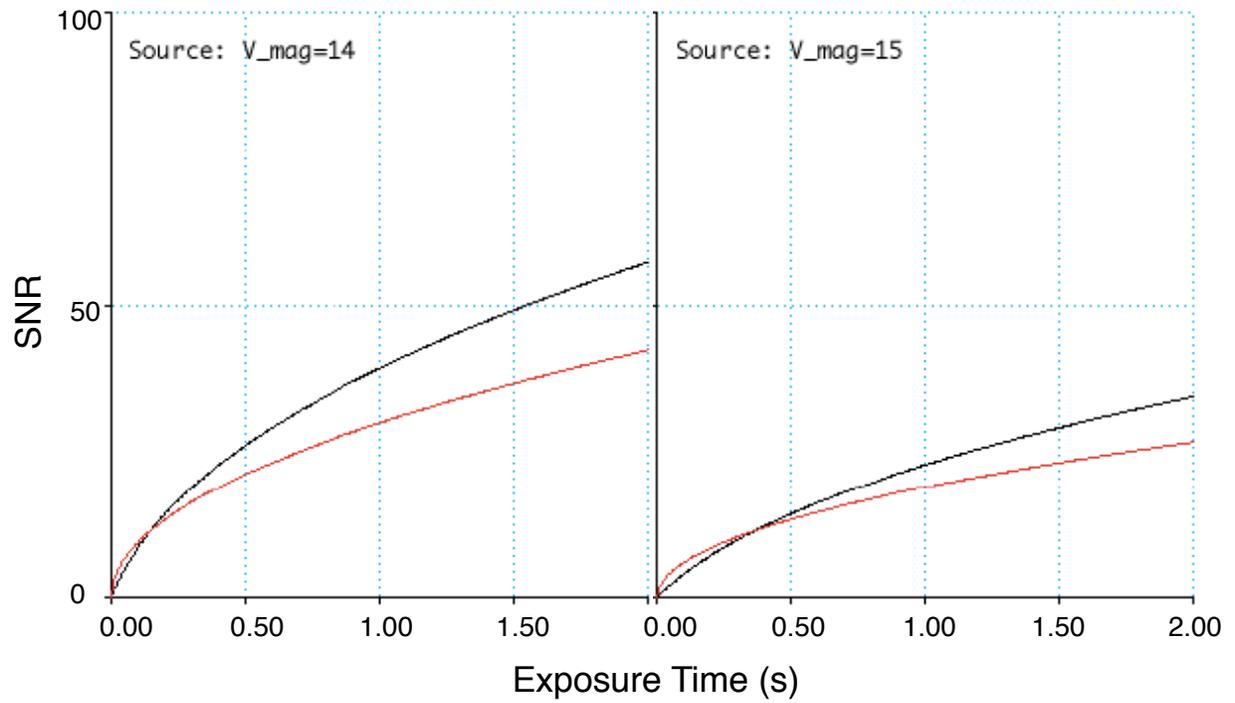

**FIG. 8. SNR Estimates for 14-Inch Meade/MicroMax 512BFT *PHOT* Systems**

SNR estimates for exposure times between 0 and 2 s for a PhotonMax 512B camera in EMCCD mode (the red lines) and a MicroMAX 512BFT camera with 3.1 e- read noise (black lines). Four source brightnesses are considered (V = 14, 15, 16 and 17), where the source brightness refers to the *combined* signal from the occulting object and the occulted star. As expected, the EMCCD has an advantage when there is not much light (i.e., for faint sources and/or short exposure times). For a V=16 source (about four times fainter than Pluto), the cross-over point is at one second. EMCCD mode is less advantageous at telescopes with apertures larger than 14 inches.

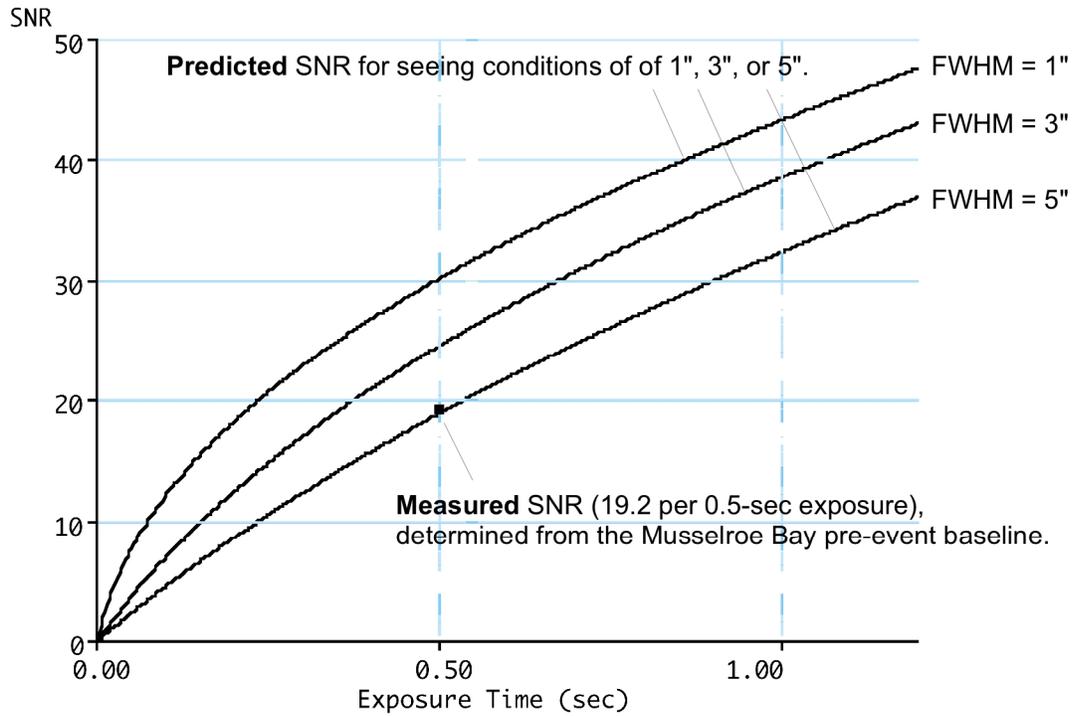

**FIG. 9. Predicted and Actual Performance of a *PHOT* System on a 14-inch Telescope**

For small telescopes and faint sources, the SNR is a sensitive function of the seeing and focus quality. This is expected: poor seeing means that point sources are spread over more pixels, resulting in a higher read noise penalty when a source's counts are summed over all those pixels. At the Musselroe Bay site (31-JUL-2007), the measured SNR was comparable to the predicted SNR under the assumption that the typical source FWHM was around 4 - 5 arcseconds, which is consistent with the image quality from that night.

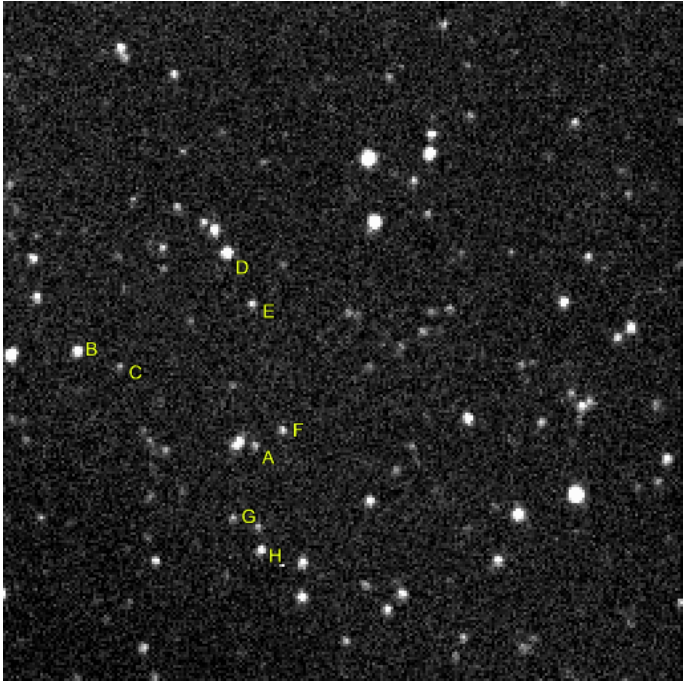

**FIG. 10. Observed Counts for Faint Stars Using *PHOT*.**

In this 0.5-sec exposure of the 12-JUL-2006 Pluto occultation field (obtained with a 14-inch Meade telescope from the Campbell Farm in Longford, Tasmania), we have extracted counts for eight faint stars to demonstrate the sensitivity of the *PHOT* cameras. The occultation star, *A*, has an estimated UCAC2 magnitude (between V and R filters) of 15.0 and generated 811 DN (data number, or counts) in this image. Given the gain (about 2 e-/DN), the background noise level (about 18 e- per pixel), the FWHM (about 2.1 pixels) and the read noise (about 3.1 e- per pixel), the measured SNR of star *A* is 2.5 in a 0.5-sec exposure. Without the high background (which roughly doubled the noise from other sources), the SNR would have been 6.4 per 0.5 s exposure.

| Star Label | DN (background-subtracted counts) per 0.5-s exposure | UCAC-Mag (from 579-642 nm) | UCAC2_ID |
|---|---|---|---|
| A | 25 | 15.0 | 26039859 |
| B | 76 | 13.2 | 26039956 |
| C | 14 | 15.2 | 26039936 |
| D | 107 | 13.5 | 26039894 |
| E | 24 | 14.2 | 26039875 |
| F | 31 | 14.3 | 26039845 |
| G | 19 | 14.7 | 26039867 |
| H | 46 | 13.7 | 26039844 |

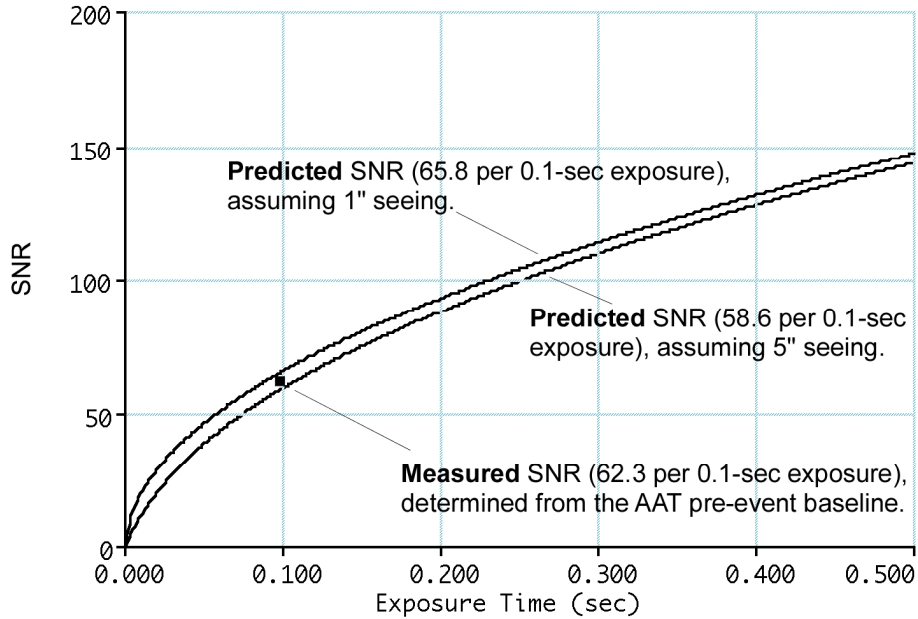

**FIG. 11. Predicted and Actual Performance of a *PHOT* System on the 3.9-m AAT Telescope**

For large telescopes, the quality of the focus or seeing barely affects the SNR of the reduced light curve. Because the read noise is a small component of the total noise, the number of pixels covered by each source only makes a small difference to the total noise estimate. In this case, the measured noise (determined from the flat, pre-event portion of the light curve) was 62.3 for each 0.1-sec timestep, close to the SNR predictions for seeing conditions between 1 - 5 arcseconds.